\documentclass[preprint,aps,prb,showpacs]{revtex4}
\usepackage{color}
\usepackage{epsfig}
\usepackage{bm}

\begin{document}
\title{Confined photon modes with triangular symmetry in hexagonal
microcavities in 2D photonic Crystals}
\author{Yuriy A. Kosevich}
\thanks{Permanent address: Semenov Institute of Chemical Physics, Russian Academy
of Sciences, ul. Kosygina 4, 119991 Moscow, Russia. yukosevich@yahoo.com.}
\affiliation{Wave Phenomena Group, Department of Electronic Engineering,
Polytechnic University of  Valencia, c/Camino de Vera s/n, E-46022 Valencia, Spain}
\author{Jos\'e S\'anchez-Dehesa}
\email[Corresponding author. E-mail:]{jsdehesa@upvnet.upv.es}
\affiliation{Wave Phenomena Group, Department of Electronic Engineering,
Polytechnic University of  Valencia,  c/Camino de Vera s/n, E-46022 Valencia, Spain}
\author{Alfonso R. Alija, Luis J. Mart\'{\i}nez, Maria L. Dotor and Pablo A. Postigo}
\affiliation{Instituto de Microelectr\'{o}nica de Madrid, Centro Nacional de Microelectr\'{o}nica, Consejo Superior de
Investigacion Cientificas, Isaac Newton 8, PTM Tres Cantos, 28760 Madrid, Spain}
\author{Dolores Golmayo}
\affiliation{Instituto de Ciencia de Materiales, Consejo Superior de Investigacion Cient\'{\i}ficas, Cantoblanco, 28049
Madrid, Spain}
\date{\today}

\begin{abstract}
We present theoretical and experimental studies of the size and thickness dependencies of the optical emission
spectra from microcavities with hexagonal shape in films of two-dimensional photonic crystal. A semiclassical
plane-wave model, which takes into account the electrodynamic properties of quasi-2D planar photonic microcavity, is developed to predict the  eigenfrequencies of the confined photon modes as a function of both the
hexagon-cavity size and the film thickness. 
Modes with two different symmetries, triangular and hexagonal, are critically analyzed.
It is shown that the model of confined photon modes with triangular symmetry gives a better agreement between the
predicted eigenmodes and the observed resonances.
\end{abstract}
\maketitle

\section{Introduction}
New physics often emerges in confined systems when reduced dimensionality leads to new symmetries that, together with
the confinement, result in new interesting phenomena. Microcavities in two-dimensional  photonic crystals (2D PhCs) are
an important example of actual confining structures. They are used for localization of photons into PhC bandgaps (BGs)
in order to build ultra-small, low-loss, low-power and low-threshold lasers and light-emitting structures \cite{noda}.
In recent years an important amount of research, both fundamental and applied, has been focused on PhCs
microcavities with hexagonal shape fabricated in semiconductor slabs perforated with holes, see, e.g.,
\cite{BookNoda,BookBenisty}, the vertical emission and lasing from which have been successfully demonstrated
\cite{painter,park,fujita}. In such microcavities, light waves having frequencies inside the photonic bandgap are trapped and develop confined photon modes, some of them can be considered as circulating around the perimeter of the cavity due to multiple total internal reflections at its boundary (whispering gallery-type modes)\cite{pottier,benisty}.
Vertical emission and lasing from such microcavities can be achieved by the coupling of the confined phonon modes (with
TE polarization) to the vacuum states.  Similar ``vertically emitting'' 2D systems are also realized in live nature,
for example in fluorescent butterfly wing scales \cite{vukusic}.

The eigenfrequencies of the confined photon modes are among most important characteristics of a microcavity: they
determine the energies of photons having maximal spontaneous emission probabilities (the Purcell effect).
In this work we present both theoretical and experimental arguments in favor of the existence of two
degenerate photon eigenmodes with $triangular$ $symmetry$ in hexagonal microcavities. A semiclassical
plane-wave model, which takes into consideration electrodynamic properties of  
quasi-2D planar photonic microcavity, is developed to predict the  eigenfrequencies of the confined photon modes as a function both of the
hexagon-cavity size and film thickness. 
We show that the eigenfrequencies
of the triangular-symmetry modes determine spectral positions of spontaneous emission peaks, which were recently
observed in photoluminescence spectra from microcavities in a 2D PhC in III-V semiconductor slabs [\onlinecite{alija}].
Here, a new set of microcavities has been built and characterized. As in Ref. [\onlinecite{alija}], by accurate
changing the PhC film thickness the wavelengths of the PhC microcavity eigenmodes are tuned in order of tens of
nanometers toward higher energies. 
A comparison of the observed eigenmodes with the spectral positions
predicted by the models of the modes with triangular symmetry and hexagonal symmetry has also been performed. From that
comparison we conclude that the proposed model of triangular-symmetry modes gives better agreement with the experiment than that based on hexagonal-symmetry modes.
 Therefore we expect that the proposed model of triangular-symmetry modes will be useful both
in understanding the physical origin and for the quantitative prediction of the spontaneous emission peaks in 2D PhC microcavities in the form of 
ideal 
and distorted hexagons.


\section{Experimental setup}
\begin{figure}[h]
\centering
\includegraphics[width=8.3cm]{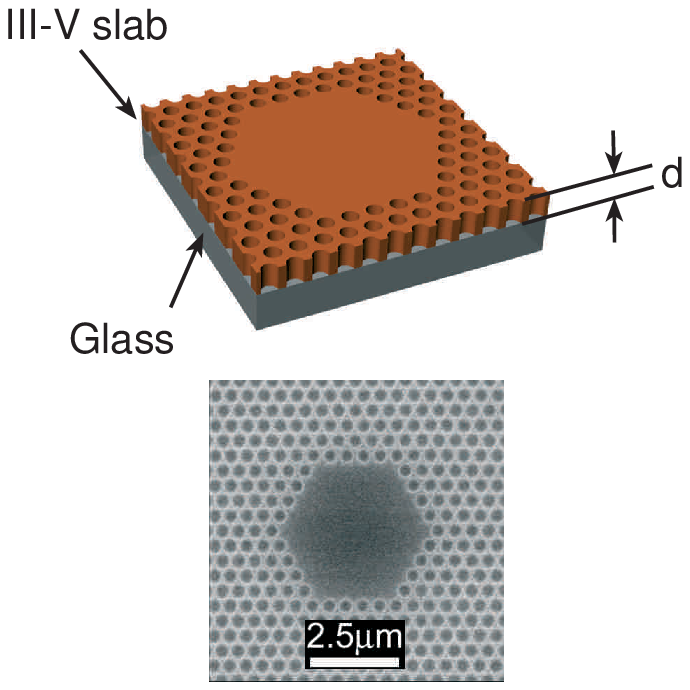}
\caption{(Lower image) Scanning electron microscopy
image of the fabricated H5 cavities. (Upper image) Schematic view of H5 microcavity fabricated on the III-V
semiconductor slab with a thickness d=265 nm bonded to a glass substrate.} 
\label{FigH5}
\end{figure}

Experimentally, several 2D PhC structures with a hexagonal lattice of circular air holes have been fabricated in an
InGaAsP semiconductor film incorporating an active medium composed of three In$_{0.47}$Ga$_{0.53}$As quantum wells.
This structure gives rise to strong photoluminescence (PL) spectra centered around 1500 nm at room temperature.
Processing of the PhC structures was done by electron-beam lithography of a poly(methyl methacrylate) (PMMA) layer on
top of a SiO$_2$ layer (120 nm thick) \cite{alija}. To provide the membranes with sufficiently strong mechanical
support, the fabricated layers were bonded to a thin borosilicate glass ($n=1.53$) with optical glue ($n=1.5$).
Circular holes were made in the InGaAsP membrane by reactive ion beam etching. A 2D hexagonal
PhC array with lattice constant $a=$ 500 nm surrounds hexagonal microcavities 
with a variable number of missing holes per side.
 Here, H3 (three missing holes per side) and H5 (three missing holes per side) microcavities 
are fabricated and
characterized. 
Figure \ref{FigH5} shows a scanning electron microscopy image of one H5 microcavity and a
descriptive drawing of the fabricated structure. 
Different values of the radius of the holes $r$ around 0.33$a$ have been used. The variation of the radius $r$ is estimated
from the scanning electron microscopy pictures, being below 5\%. With this choice of parameters, a large band gap
exists for the TE mode between the normalized frequencies of $\omega a/(2\pi c)\sim 0.28$  and $\omega a/(2\pi c)\sim
0.36$, which corresponds to the wavelength range of 1250-1600 nm. The fabricated structures are optically pumped with a
780 nm laser diode. An objective lens (0.40 NA) is used to focus the excitation spot. The size of the excitation spot
was around 3.5 $\mu$m which is small enough to fit inside the PC structures and to generate cavity modes. Light is
collected by a lens inside a 0.22 m monochromator with a cooled InGaAs photodiode connected to a lock-in amplifier. The
resolution of the experimental setup in this configuration is around 2.5 nm.


\section{Model}
\begin{figure} [h] 
\includegraphics[width=85mm]{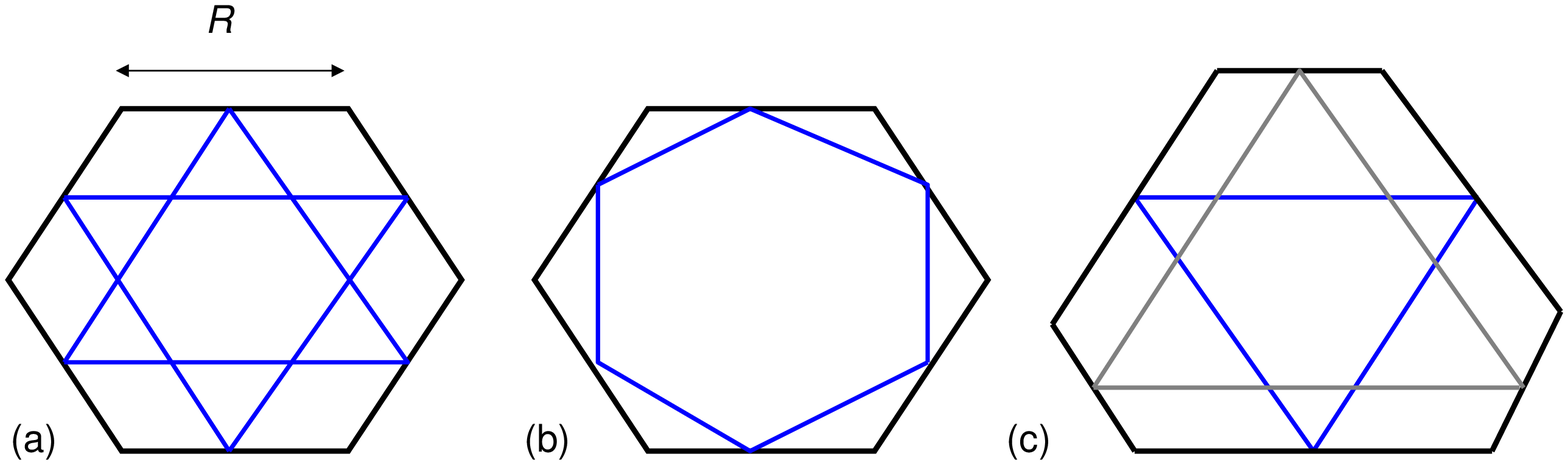}
\caption{(color online) Ray paths in triangular-symmetry modes in ideal, (a), and distorted, (c), hexagons; (b) presents ray path for the hexagonal-symmetry mode in an ideal hexagon. Two triangular-symmetry modes are degenerate in an ideal hexagon, and have different optical paths and eigenfrequencies in a distorted hexagon.}
\label{fig2}
\end{figure}

Now we turn to the description of our model. Figure \ref{fig2}(a) shows the hexagonal cavity and the ray paths, which
belong to the two degenerate triangular-symmetry modes. 
Within the semiclassical plane-wave model, the dispersion equation for the degenerate triangular-symmetry photon modes, confined in an ideal hexagonal microcavity in 2D PhC,  can be obtained  from the requirement that the total phase shift of the wave along its closed path is an integer multiple of $2\pi$:
\begin{eqnarray}
\label{eq1}
k_{i}^{(N)}4.5 R - 3\phi&=&2\pi N, \\
\frac{\omega_{N}}{2\pi}=\frac{c_{i}}{\lambda_{i}^{(N)}}&=&\frac{c}{4.5Rn_{eff}}(N+\frac{3\phi}{2\pi}),
\label{eq2}
\end{eqnarray}
where $N$ is integer ``quantum'' wavenumber, $\phi$ takes into account the additional phase shift that occurs during the total internal reflection of the confined photon mode in the 2D PhC bandgap frequency region, $c_{i}$ and $k_{i}=2\pi/\lambda_{i}$ are, respectively, the wave velocity and the $in$-$plane$ wave-vector inside the cavity. It will be shown below that the corresponding wavenumbers $N$ in the considered H3 and H5 microcavities are rather large, $N\gg 1$, which justifies the use of the semiclassical plane-wave model in resolving the eigenfrequencies.

We assume that in (or close to) the middle of the PhC bandgap the additional phase shift $\phi$ in Eqs. (1) and (2) is equal to the Keller additional phase shift due to leakage of (electron) wavefunction into classically forbidden regions, $\phi=\frac{\pi}{2}$, see Refs. [\onlinecite{keller}]. In the middle of the PhC BG, the phase shift $3\phi$ does not depend on the exact form and steepness of the effective confining potential, is independent of wavenumber $N$ and depends only on the number of turning points (equal to 3 or 6 for the triangular- or hexagonal-symmetry modes, respectively, see Figs. \ref{fig2}(a) and \ref{fig2}(b)). It is worth mentioning that we assume that in the middle of the BG the additional phase shift $\phi=\frac{\pi}{2}$
in Eqs. (1) and (2) is fixed and
does not explicitly depend on
the phase of the coefficient of the total internal reflection of the confined photon modes, in contrast to the similar to Eqs. (1) and (2) semiclassical plane-wave
 model which was applied  to the total-internal-reflection dielectric microresonators with hexagonal cross section, see Ref. [\onlinecite{nobis}]. The point is that the phase of the coefficient of the total internal reflection is different (and the difference is equal to $\pi$) for the reflection coefficients of the ${\bf E}$ or ${\bf H}$ fields, which results in different signs of the normal component of the Pointing vector, ${\bf P}=\frac{c}{4\pi}{\bf E}{\bf \times}{\bf H}$, for the incident and reflected electromagnetic waves,
see, e.g.,  Ref. [\onlinecite{landau}]. Therefore with our choice of $\phi$ in Eqs. (1) and (2), the wavenumber $N$ is uniquely defined, as in the case of semiclassical Bohr-Sommerfeld quantization condition for the bound electron motion.
If we apply the same model to the hexagonal-symmetry photon mode, confined in 2D PhC hexagonal
microcavity (in the middle of the PhC BG), see Fig. \ref{fig2}(b), $4.5 R$ and $3\phi$ in l.h.s. of Eq. (1) and r.h.s. of Eq. (2) should be replaced, respectively,  by $3\sqrt{3}R$ and $6\phi$, with $\phi=\frac{\pi}{2}$.

The degeneracy of the two triangular-symmetry modes in an ideal hexagonal microcavity can be removed in a ``distorted hexagon'', which is bound by two regular triangles with different sides (see Fig. \ref{fig2}(c)). In this case the total optical paths and correspondingly the eigenfrequencies of the two modes, bound inside the different triangles, will be different.

The cavity eigenfrequencies $\omega_{N}$ in Eq. (2) are determined by the effective refractive index of the system $n_{eff}\equiv c/c_{i}$. The $n_{eff}$ is determined in turn by electrodynamic properties of 
quasi-2D planar photonic microcavity. 
The approach of the effective refractive index allows us 
to describe the actual 3D system as an effective 2D system,  
see also Ref. [\onlinecite{qiu}]. The $n_{eff}$ depends on the polarization 
of the confined photon mode (TE or TM), and on the relative film thickness 
$k_{i}d/2\pi=d/\lambda_{i}$ of the quasi-2D planar structure. 
It is possible to obtain an explicit dependence $n_{eff}$$=$$n_{eff}(d/\lambda_{i})$ by solving dispersion equation for electromagnetic mode in a waveguide formed by a dielectric film
with thickness $d$ and dielectric constant $\epsilon_{f}$, sandwiched between a substrate and a top with dielectric constants $\epsilon_{s}$ and  $\epsilon_{t}$, respectively, see, e.g., Ref.  [\onlinecite{haus}].
The dispersion equation for the angular frequency $\omega$ of TE mode with a given in-plane wave vector $k_{i}$ in such planar waveguide system
has the following form \cite{haus}:
\begin{eqnarray}
\label{eq3}
\tan\kappa_{f}d&=&\frac{(\kappa_{t}+\kappa_{s})\kappa_{f}}{\kappa_{f}^{2}-\kappa_{t}\kappa_{s}}, \\
\kappa_{f}&=&\sqrt{\frac{\omega^{2}}{c^2}\epsilon_{f}-k_{i}^{2}},~~ \kappa_{t,s}=\sqrt{k_{i}^{2}-\frac{\omega^{2}}{c^2}\epsilon_{t,s}} \nonumber.
\end{eqnarray}
Equation (3) can be cast in the following transcendent equation which relates $\epsilon_{eff}$=$n_{eff}^2$, $\epsilon_{f}$$>$$\epsilon_{eff}$$>$($\epsilon_{t}$,$\epsilon_{s}$),  with the dimensionless in-plane wavenumber  $k_{i}d$:
\begin{eqnarray}
&&\tan\left[k_{i}d\sqrt{\frac{\epsilon_{f}}{\epsilon_{eff}}-1}\right]= \nonumber \\
&&\frac{\left[\sqrt{1-\frac{\epsilon_{s}}{\epsilon_{eff}}}+\sqrt{1-\frac{\epsilon_{t}}{\epsilon_{eff}}}\right]\sqrt{\frac{\epsilon_{f}}{\epsilon_{eff}}-1}}{\frac{\epsilon_{f}}{\epsilon_{eff}}-1-\sqrt{1-\frac{\epsilon_{s}}{\epsilon_{eff}}}\sqrt{1-\frac{\epsilon_{t}}{\epsilon_{eff}}}}.
\label{eq4}
\end{eqnarray}

\begin{figure}
\centering
\includegraphics[width=6.5cm]{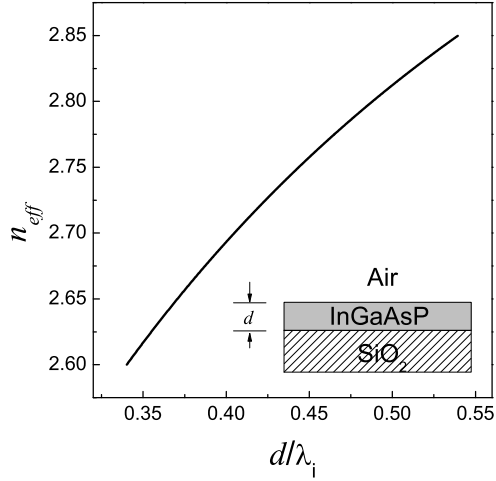}
\caption{Effective refractive index $n_{eff}$ versus InGaAsP layer thickness over internal wavelength
$d/\lambda_{i}$  in a planar waveguide with  $\epsilon_{f}=10.89$ (InGaAsP), $\epsilon_{s}=2.34$ (SiO$_2$) and
$\epsilon_{t}=1$ (air).} 
\label{fig3}
\end{figure}

Figure \ref{fig3} presents a numerical solution of Eq. (4) for the TE$_{0}$ mode for $n_{eff}$ as a function of
$d/\lambda_{i}$ (in the interval of interest for our microcavities) in a InGaAsP planar waveguide bonded on a SiO$_2$
substrate, and using air as top layer. Figure \ref{fig3} shows that $n_{eff}$ is a monotonously increasing function of
$d/\lambda_{i}$, which can be explained by the stronger confinement inside the high-index InP layer of the cavity modes
with the smaller, with respect to the layer thickness, wavelengths. Qualitatively similar dependence of  $n_{eff}$ on
$d$ in a 2D PhC system was previously obtained in Ref. [\onlinecite{lee}] by numerical evaluation of the space-averaged
electric-field mode energy.

\section{Results and discussion}

\begin{figure} [ht] 
\centering
\includegraphics[width=8.3cm]{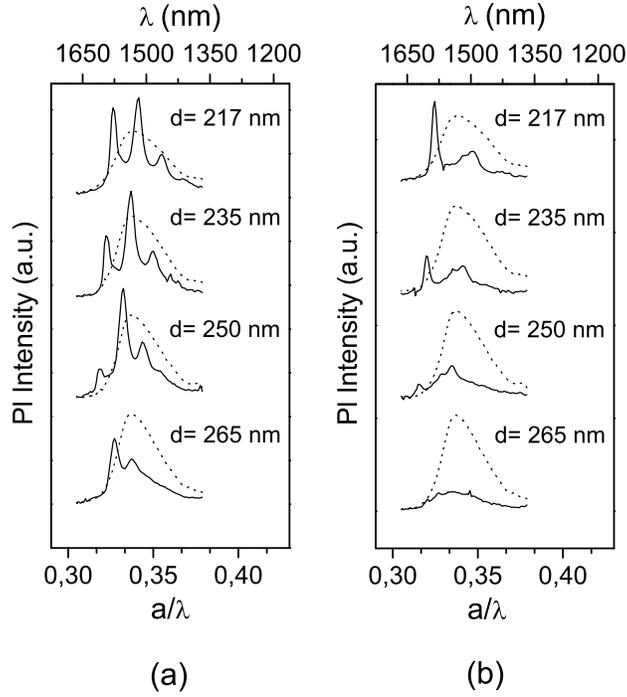}
\caption{Photoluminescence spectra of the H5 (a) and H3 (b) microcavities
with different slab thicknesses $d$ (solid lines). The spectra of an unpatterned region in the vicinity of the microcavity is
also shown (dotted lines). Here $\lambda$ is the vacuum wavelength: $a/\lambda$=$\omega a/2\pi c$. }
\label{figH5_H3}
\end{figure}

\begin{figure}[ht]
\centering
\includegraphics[width=8.3cm]{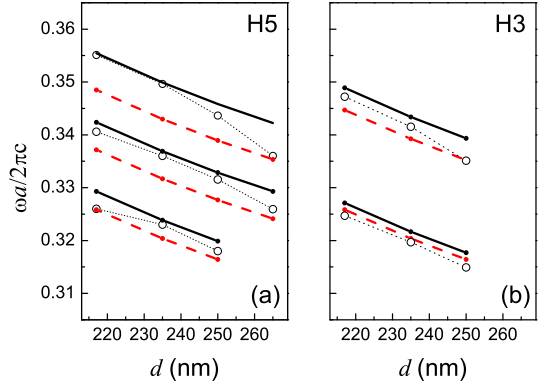}
\caption{(color online) Normalized frequencies of emission peaks produced by localized modes in H5 (a) and H3 (b) microcavities as a function of InP layer thickness, $d$. Open circles represent the experimental data obtained by microphotoluminescence. The dotted lines are guides for the eye. Solid (red dashed) lines show the theoretical prediction given by the model of confined triangular-symmetry (hexagonal-symmetry) modes.}
\label{fig5}
\end{figure}

The $d/\lambda_{i}$-dependence of $n_{eff}$ in Eq. (2) furnishes the clue to a quantitative description within the proposed model
of the observed $d$-dependence of the microcavity eigenfrequencies. To describe the predicted eigenfrequencies
(emission peaks) in the measured optical spectra, see Fig. \ref{figH5_H3}, 
we use $R=5a$ and $R=3a$ as the side lengths of the H5 and H3 microcavities in Eqs. (1) and (2).
Symbols in Fig. \ref{fig5} show the spectral positions of the observed absolute eigenfrequencies (in reduced units),
$\omega_N a/2\pi c$, as a function of the InGaAsP film thickness ($d$), while solid (dashed) lines give the predicted
by Eqs.\,(1) - (4) eigenfrequencies of the confined triangular-symmetry (hexagonal-symmetry) photon modes, without any
fitting parameters.

Within the model of triangular-symmetry modes, $N=19$, $N=20$ and $N=21$ correspond, respectively,  to the 1st, 2nd and 3rd peak in the H5 cavity, and  $N=11$ and $N=12$
correspond to the 1st and 2nd peak in the H3 cavity. (Within the model of hexagonal-symmetry modes, $N_{hex}=21$, $N_{hex}=22$ and $N_{hex}=23$ correspond, respectively,  to the 1st, 2nd and 3rd peak in the H5 cavity, and  $N_{hex}=11$ and $N_{hex}=12$
correspond to the 1st and 2nd peak in the H3 cavity.) The corresponding wavenumbers $N$ (and $N_{hex}$) are indeed large, as it was
mentioned above. The best agreement of the model of triangular-symmetry modes with the measurements is reached for the half-wavelength layer, $d=$235 nm$\approx
\lambda/2n_{f}$, $\lambda$ being the vacuum wavelength. In this case the wave path is indeed close to the
two-dimensional path in the plane of the microcavity, while the path becomes more three-dimensional for the larger slab
thicknesses. It is seen in Fig. \ref{fig5} that the model of triangular-symmetry modes describes better the absolute eigenfrequencies of the observed confined modes both in H5 and H3 microcavities. 

\begin{figure} [ht] 
\centering
\includegraphics[width=8.3cm]{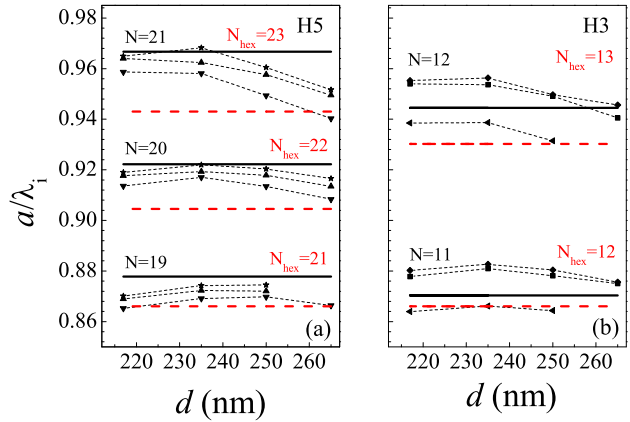}
\caption{ (color online) Normalized eigenfrequencies with corresponding wavenumbers $N$ and $N_{hex}$ (for the triangular- and hexagonal-symmetry modes,  respectively) 
of emission peaks produced by confined photon modes in H5 (a) and H3 (b) microcavities as a function of InP layer thickness 
$d$. Different bold symbols represent the experimental data obtained by microphotoluminescence.
Solid (red dashed) lines
represent the theoretical prediction given by the model of confined triangular-symmetry (hexagonal-symmetry) modes.}
\label{fig6}
\end{figure}

Figure \ref{fig6} shows dimensionless eigenfrequencies $\omega an_{eff}/2\pi c$=$a/\lambda_{i}$ with corresponding wavenumbers $N$ and $N_{hex}$ (for the triangular- and hexagonal-symmetry modes, respectively) versus  InGaAsP layer
thickness. Now, solid (dashed) lines represent the  constant values predicted by Eq. (2) for the confined
triangular-symmetry (hexagonal-symmetry) modes, and different bold symbols define the peaks in the optical spectra (not shown) measured  in a
new set of H5 and H3 fabricated microcavities. To include the experimental data in Fig. \ref{fig6}, we have used Eq.
(4) [for the same $n_{eff}(d/\lambda_{i}^{(N)})$ as in Fig. \ref{fig5}]. 

The agreement found with the new data also supports our model of confined photon modes with triangular symmetry. 
Another evidence in favor of the proposed semiclassical plane-wave model
can be obtained 
from previously published measurements of photoluminescence spectra from 
hexagonal microcavities in 2D photonic crystlas. For instance, one can deduce 
from the spectra in Ref. [\onlinecite{monat}] that the spacing between 
the strongest photoluminescence peaks in H2 and H5 hexagonal microcavities 
in 2D PhCs scales with the cavity size R as 1/R. Namely, the spacing between the
two strongest neighbor peaks seen in Figs. 2b and 2c of Ref. [\onlinecite{monat}] are  
approximately equal to 130 and 53 nm for H2 and H5 microcavities, respectively.
 The ratio of these peak spacings is 
indeed close to the value 5/2, given by the inverse ratio of the cavity sizes. 
Another earlier observation, to which we can refer in support of our 
proposed  plane-wave model is the blue shift of spontaneous emission spectra from the 
suspended 2D PhC microcavities, surrounded from both sides by air, with respect 
to the spectra from microcavities grown on Si wafer with 
SiO$_{2}$ transfer layer, see Figs. 1 and 8 in Ref. [\onlinecite{monat}]. 
The effective refractive index  $n_{eff}$ of the planar photonic microcavity 
is lower in the former structure 
which results in the blue shift of the cavity 
eigenfrequencies, in accordance with Eq. (2). 

In connection with aforementioned comparison between the predicted and observed microcavity photon 
eigenmodes, it is important to emphasize that the reason why the triangular-symmetry modes give   
the predominant contribution to photoluminescence spectra in actual heterostructures is possibly 
related with relatively high $modal$ $volume$ of these modes which increases the coupling of these modes 
to the vacuum states, see, e.g., Ref. [\onlinecite{benisty}]. 
Namely, as follows from the comparison of the ray paths for the triangular- and hexagonal-symmerty modes 
in Figs. 2(a) and 2(b), the hexagonal-symmetry modes are confined more close 
to the cavity boundary and therefore give less contribution to the far-field emission from the cavity, 
which comes mainly from the "bulk-like" cavity eigenmodes. The same arguments can be applied to the 
comparison between the emission strengths of the bulk-like trianguar-symmetry modes in hexagonal microcavities 
and a dense set of cavity 
eigenmodes, which were predicted for H5 hexagonal microcavities in 2D PhC with the use of finite-difference 
time-domain simulations in Ref. [\onlinecite{benisty}]. As it is explained in  Ref. [\onlinecite{benisty}], 
only few of the predicted eigenmodes can emit strong enough far-field radiation, the other eigenmodes have very 
"spotty patterns" which suppress the emission.           
It is also worth mentioning in this connection that the possible origin of some of the less strong peaks, 
in comparison with that given by the triangular-symmetry modes,  
in the observed photoluminescence spectra [Fig. 2 in Ref. [\onlinecite{monat}]] can be the additional cavity eigenmodes 
caused by the splitting of the degenerate 
triangular-symmetry eigenmodes in hexagonal cavities by fabrication imperfections and defects. The splitting of spectrally 
degenerate cavity modes by unavoidable cavity imperfections 
was previously observed and studied, see, e.g., Refs. [\onlinecite{painter1,hennesy}].

At this point, let us remember that dielectric microresonators and 
microcavities with hexagonal shape, which are often
called as whispering-gallery resonators, have also attracted much interest 
for possible optical applications in recent
years. The whispering-gallery modes in dielectric resonators with hexagonal shape correspond to 
the hexagonal-symmetry modes in hexagonal
microcavities in 2D PhC (see Fig. \ref{fig2}(b)).  The ``horizontal'' emission 
from such dielectric microresonators is
achieved by the coupling of the confined phonon modes 
with TM polarization to the vacuum states. A semiclassical plane-wave model of triangular-symmetry modes
can also be applied to quasi-2D planar dielectric microresonators. Recent studies of
dielectric microresonators made of zinc oxide (ZnO) nanoneedles with hexagonal 
cross section have been reported in Ref. 
\onlinecite{nobis}. The refractive index n 
 of ZnO is close to 2 and, therefore, the
critical angle of incidence for total internal reflection in the microresonator is close to $\pi/6$. 
Since the angle of incidence of
triangular-symmetry (whispering-gallery) modes in hexagonal microresonator is $\pi/6$ ($\pi/3$), see Figs.
\ref{fig2}(a) and \ref{fig2}(b), the triangular-symmetry modes in planar, 
with half-wavelength thickness $d\approx
\lambda/2n$, dielectric microresonators made of ZnO can be considered as
$radiating$ counterparts of $non$-$radiating$ whispering-gallery modes.

\section{Summary}

In summary, theoretical and experimental evidences have been presented showing that confined double-degenerated photon eigenmodes with triangular symmetry can be the origin of the
emission peaks observed in the optical spectra from planar hexagonal microcavities in 2D photonic crystals. 
Here, an analytical semiclassical plane-wave model, which  takes into consideration electrodynamic properties of  
quasi-2D planar photonic microcavity, has been developed for the quantitative evaluation of the eigenfrequencies of the photon modes, laterally localized due to the total internal reflection in the middle of photonic crystal bandgap. 
This analytical model describes quantitatively the experimentally observed cavity-size and photonic-crystal-thickness dependencies of the eigenfrequencies of the confined photon modes.
Removal of the degeneracy and change of eigenfrequencies of the confined photon modes are predicted for hexagonal 
microcavities with imperfections and for microcavities in the form of distorted hexagon.

\section{Acknowledgements}
Work supported by Ministry of Science and Education (MEC) of Spain (Refs. TEC2004-03545, TEC2005-05781-C03-01,
NAN2004-08843-C05-04, NAN2004-09109-C04-01), and contracts S-505/ESP/000200, UE NoEs SANDIE (NMP4-CT-2004-500101) and
PHOREMOST (IST-2-511616-NOE). The authors acknowledge useful discussions with Andreas H\aa kansson, Javier Mart\'i and
Daniel Torrent. Yu. A. K. acknowledges a support from MEC (Grant SAB2004-0166). A. R. A. thanks a FPU fellowship (Ref.
AP2002-0474) and L. J. M. an I3P fellowship.

\newpage

\end{document}